\documentclass[lettersize,journal]{IEEEtran}
\usepackage{amsmath,amsfonts}
\usepackage{algorithmic}
\usepackage{algorithm}
\usepackage{array}
\usepackage[caption=false,font=normalsize,labelfont=sf,textfont=sf]{subfig}
\usepackage{textcomp}
\usepackage{stfloats}
\usepackage{url}
\usepackage{verbatim}
\usepackage{graphicx}
\usepackage{cite}
\begin{document}

\title{Dual-layer Image Compression via Adaptive Downsampling and Spatially Varying Upconversion}

\author{Xi~Zhang and
        Xiaolin Wu,~\IEEEmembership{Fellow,~IEEE}
        % <-this % stops a space
% \thanks{This work was supported in part by the National Natural Science Foundation of China (NSFC) and Natural Sciences and Engineering Research Council of Canada (NSERC).}
% <-this % stops a space
\thanks{X.~Zhang is with the Department of Electronic Engineering, Shanghai Jiao Tong University, Shanghai, China (email: zhangxi\_19930818@sjtu.edu.cn).}
% <-this % stops a space
\thanks{X.~Wu is with the Department of Electrical \& Computer Engineering, McMaster University, Hamilton, L8G 4K1, Ontario, Canada (email: xwu@ece.mcmaster.ca).}}

\markboth{Journal of \LaTeX\ Class Files,~Vol.~14, No.~8, August~2021}%
{Shell \MakeLowercase{\textit{et al.}}: A Sample Article Using IEEEtran.cls for IEEE Journals}


\maketitle

%%%%%%%%% ABSTRACT
\begin{abstract}
  Ultra high resolution (UHR) images are almost always downsampled to fit small displays of mobile end devices and upsampled to its original resolution when exhibited on very high-resolution displays. This observation motivates us on jointly optimizing operation pairs of downsampling and upsampling that are spatially adaptive to image contents for maximal rate-distortion performance. In this paper, we propose an adaptive downsampled dual-layer (ADDL) image compression system. In the ADDL compression system, an image is reduced in resolution by learned content-adaptive downsampling kernels and compressed to form a coded base layer.  For decompression the base layer is decoded and upconverted to the original resolution using a deep upsampling neural network, aided by the prior knowledge of the learned adaptive downsampling kernels. We restrict the downsampling kernels to the form of Gabor filters in order to reduce the complexity of filter optimization and also reduce the amount of side information needed by the decoder for adaptive upsampling.
  Extensive experiments demonstrate that the proposed ADDL compression approach of jointly optimized, spatially adaptive downsampling and upconversion outperforms the state of the art image compression methods.
\end{abstract}

%%%%%%%%% BODY TEXT
\section{Introduction}
\label{sec:intro}
Over the past decade, as smartphone cameras started to rival and even beat DSLR cameras in resolution, a large number of
ultra high resolution (UHR) images, each of more than ten million pixels, are produced, stored and transmitted everyday. However, these images are almost always down sampled from its original resolution to fit small displays of mobile end devices.  Also in web applications, users often want to browse through a set of images quickly in relatively low resolution to save time and communication bandwidth.  All these practices raise a question, why not keep
UHR images in a downsampled version for routine uses, and only upsample to original ultra high resolution when needed, say to be exhibited on very high-resolution displays?  Such a two-layer image representation, in our opinion, improves both the operability and bandwidth economy.  One can appreciate the savings in storage and bandwidth for internet content providers if UHR images are coded in this new presentation.
These observations motivate our research on jointly optimizing operation pairs of downsampling and upsampling that are spatially adaptive to image contents for maximal rate-distortion performance.

Image downsampling is one of the most common image processing operations, aiming to reduce the resolution of the high-resolution (HR) image while retaining the maximum amount of information.
%keeping its visual appearance.
According to the Nyquist-Shannon sampling theorem~\cite{shannon1949communication}, high-frequency contents will inevitably get lost after downsampling.
%during the downsampling process.
Opposite to image downsampling is image upsampling, also known as super-resolution (SR), with the goal of recovering the underlying HR image from the given LR input.
Image SR is an ill-posed inverse problem because an undersampled image can be the result of down sampling many different HR images.  Due to this uncertainty, the
image SR methods are bound to erase or distort high-frequency 
features~\cite{yang2017image,yang2010image,dong2011image,dong2010super}.
%The quality of the SR result is very limited due to the ill-posed nature of the problem and the lost high-frequency components are hard to be well-recovered.

%Image upsampling techniques pay much attention to enhance details, such as edges, which helps to improve human visual perception.
%On the other hand, recent state-of-the-art deep SR models have witnessed their capability to restore HR image from the
%LR version downsampled using traditional filtering-decimation
%based methods with great performance gain.
%However, the predetermined downsampling operations may be sub-optimal to the SR task and state-of-the-art deep SR models still cannot well
%recover fine details from distorted textures caused by the fixed downsampling operations.

%Previous works regard image upsampling and SR as separate tasks.

Previously, image downsampling and image SR are studied as separate problems.
In this paper, we jointly design image down- and up-sampling operators and propose a new methodology of adaptive downsampled dual-layer image compression (ADDL).
The paired down- and up-sampling tasks are fulfilled by neural networks.
In the proposed ADDL image compression system, an image is downsampled by a bank of learned content-adaptive downsampling kernels and then compressed to get the low-resolution (LR) layer.  The LR layer can be decompressed and used as is, and it can also be upconverted, if so wished, to the original high resolution using a deep upsampling neural network. The latter task is aided by the prior knowledge of the learned adaptive downsampling kernels. Unlike the existing method~\cite{sun2020learned} that directly optimizes the weights of the downsampling kernels using a deep neural network, we restrict the downsampling kernels to the form of Gabor filters and optimize the filter parameters at each pixel location. This allows us to greatly reduce the complexity of the downsampling kernel optimization network.

In addition, downsampling kernels contain the information of the high-frequency content which is lost in the downsampling process, such as the orientation of the edges, so they are useful for reconstructing the high-frequency textures in the upconversion.  However, transmitting the downsampling kernels costs extra bits and hence reduces coding efficiency, even though we only need to transmit the parameters of the downsampling Gabor filters.
Instead, we propose to predict the learned Gabor filter parameters from the compressed low-resolution layer using a lightweight network, and then further quantize the prediction residues for transmission.

In synchronization, the decoder uses the same network to predict the downsampling filter parameters from the received low-resolution layer and add the quantized residues back to estimate the Gabor filter parameters used by the encoder.
To optimize the upsampling process, we design an upsampling network which incorporates the compressed low-resolution layer and the content-adaptive Gabor filter parameters to reconstruct the high-resolution image.
Specifically, we modulate the standard convolution by Gabor filter parameters to obtain spatially variant convolution and build the upsampling network using the so-called Gabor-filter-induced spatially adaptive convolution (GSAC).

A highly desirable property of the ADDL image compression strategy is its scalability, which is important to omnipresent wireless visual communications practised at homes and offices.
In wireless networks the bandwidth is always at a premium, and end devices have diverse display capabilities,
ranging from small screens of cell phones to regular screens of laptops,
and even to very large displays and projection screens.
In such heterogeneous wireless environments, existing scalable or layered
image compression methods (e.g., JPEG 2000) are less inefficient than ADDL,
because the refinement portion of the scalable code stream still
consumes significant bandwidth and yet generates no benefits for low-resolution devices.

Furthermore, because the down-sampled image is only a small fraction
of the original size, ADDL greatly reduces the encoder complexity,
regardless what third-party codec is used in conjunction.
This property allows the system to shift the computation
burdens from the encoder to decoder, making ADDL an attractive
asymmetric compression solution when the encoder is resource
deprived.
% for instance, a small mobile device of limited computing power and tight energy budget.

The paper is organized as follows. Section~\ref{sec:related} describes the related works.
Section~\ref{sec:addl} presents the main contribution of this paper:
the design of the proposed ADDL image compression system.
We report and discuss the experimental results in Section~\ref{sec:exps}.
Section~\ref{sec:Conclusion} concludes the paper.

\section{Related Work}
\label{sec:related}

\subsection{Image super resolution and rescaling}
Image super resolution aims to reconstruct the underlying high-resolution image given the downsampled Low-resolution image. After SRCNN~\cite{dong2014srcnn}, the first CNN-based super-resolution method, many other CNN-based models have been proposed in recent years~\cite{dong2014srcnn,kim2016vdsr,ledig2017SRGAN,wang2018esrgan,zhang2018rcan,zhang2018RDN,liu2020RFANet,zhang2020usrnet,liang21swinir,kai2021bsrgan,liang21manet,zhang2021DPIR}.
However, these newly proposed methods tend to produce over-smoothed results when trained with the pixel loss.
To solve this problem, the perceptual loss~\cite{johnson2016perceptual,wang2018esrgan} and GAN loss~\cite{goodfellow2014generative,ledig2017SRGAN,wang2018esrgan,zhang2019ranksrgan} are proposed to enhance the details of the generated reuslts and improve the perceptual quality. 
% Although these progresses are highly appreciated, they are still struggling with learning a deterministic mapping from LR to HR space, which is problematic for image super resolution since single LR image may correspond to multiple HR images.

A related problem is image rescaling. It is to 
downsample the HR image to a visually meaningful LR image, 
while facilitating scale up to original HR image.
Different from image super resolution that works on a given downsampling scheme (e.g. bicubic downsampling), 
image rescaling tries to retain as much information in the LR image as possible for a better subsequent HR reconstruction. 
Image rescaling is mainly used to support resolution conversions between large and low resolution displays.
% Image rescaling can be useful for the purpose of reducing the storage and bandwidth cost. 
% In other words, it can define its own LR image space which is expected to be more informative than that by simple downsampling such as bicubic downsampling. 
In general, in image rescaling, the downsampling and upsampling processes are jointly modelled  and optimized by an encoder-decoder framework~\cite{kim2018task,li2018learning,sun2020learned,xiao2020IRN}, so that the downsampling model is optimized for the later upsampling operation.
% Recently, IRN~\cite{xiao2020IRN} proposes to use a bijective invertible neural network to model the downsampling and upsampling processes. High-frequency component is well-captured and transformed to a structured latent space in training. In testing, the HR image can be recovered by inputting the generated LR image and a randomly sampled latent variable.

\subsection{Image compression artifacts reduction}
There is a large body of literature on removing compression artifacts in images~\cite{rw_foi,rw_zhang,rw_li,rw_chang,rw_dar,rw_liu,zhou2011,zhou2012,shu2017,calic,davd,mmsd,mdvd}. The majority of the studies on the subject focus on post-processing JPEG images to alleviate compression noises, apparently because JPEG is the most widely used lossy compression standard.
Inspired by successes of deep learning in image restoration, a number of CNN-based compression artifacts removal methods were developed~\cite{ARCNN,rw_svoboda,CAR_guo,CAR_galteri}.
Borrowing the CNN designs for super-resolution (SRCNN), Dong~\textit{et~al.}~\cite{ARCNN} proposed an artifact reduction CNN (ARCNN). The ARCNN has a three-layer structure: a feature extraction layer, a feature enhancement layer, and a reconstruction layer. This CNN structure is designed in the principle of sparse coding.  It was improved by Svoboda~\textit{et~al.}~\cite{rw_svoboda} who combined residual learning and symmetric weight initialization.
Guo~\textit{et~al.}~\cite{CAR_guo} and Galteri~\textit{et~al.}~\cite{CAR_galteri} proposed to reduce compression artifacts by Generative Adversarial Network (GAN), as GAN is able to generate sharper image details.
Zhang~\textit{et~al.}~\cite{ultra} proposed to incorporate an $\ell_\infty$ fidelity criterion in the design of networks to protect small, distinctive structures in the framework of near-lossless image compression.
Mukati~\textit{et~al.}~\cite{mukati2022deep} proposed a novel $\ell_\infty$-constrained light-field image compression
system that has a very low-complexity DPCM encoder and a
CNN-based deep decoder. 
% Targeting high-fidelity reconstruction,
% the CNN decoder capitalizes on the $\ell_\infty$-constraint and light
% field properties to remove the compression artifacts and achieves
% significantly better performance than existing state-of-the-art $\ell_2$-
% based light field compression methods.

\subsection{Learning based image compression}
Much progress has been made on learning based image compression after the pioneering work of Toderici~\textit{et~al.} \cite{toderici2015} to exploit recurrent neural networks for learned image compression.
To make the network end-to-end trainable, 
the non-differential quantization is shown to be approximated by a differentiable process,
so is context modeling in entropy coding~\cite{balle2016,theis2017,agustsson2017}.
After that, a number of methods focusing on the network design are proposed.
% Toderici~\textit{et~al.} \cite{toderici2017} used recurrent neural networks (RNNs) to compress the residual information recursively.
Johnston~\textit{et~al.} \cite{johnston2018} published a spatially adaptive bit allocation algorithm that efficiently uses a limited number of bits to encode visually complex image regions.
Rippel~\textit{et~al.} \cite{rippel2017,agustsson2019} proposed to learn the distribution of images using adversarial training to achieve better perceptual quality at extremely low bit rate.
Li~\textit{et~al.} \cite{li2018} developed a method to allocate the content-aware bit rate under the guidance of a content-weighted importance map.
Some recent papers focused on investigating the adaptive context model for entropy estimation to achieve a better trade-off between reconstruction errors and required bits (entropy)~\cite{mentzer2018,balle2018,minnen2018,nonlinear,lee2018}, among which
the CNN methods of \cite{minnen2018,lee2018} are the first to outperform BPG in PSNR.
Choi~\textit{et~al.} \cite{choi2019} published a novel variable-rate learned image compression framework with a conditional auto-encoder.
Cheng~\textit{et~al.} \cite{cheng2020} proposed to use discretized Gaussian Mixture Likelihoods to parameterize the distributions of latent codes and achieved a more accurate and flexible entropy model.
Zhang~\textit{et~al.} \cite{agdl} proposed a deep learning system for attention-guided
dual-layer image compression (AGDL) by introducing a novel idea of critical pixel set.

\begin{figure*}[t]
	\centering
	\includegraphics[width=0.99\textwidth]{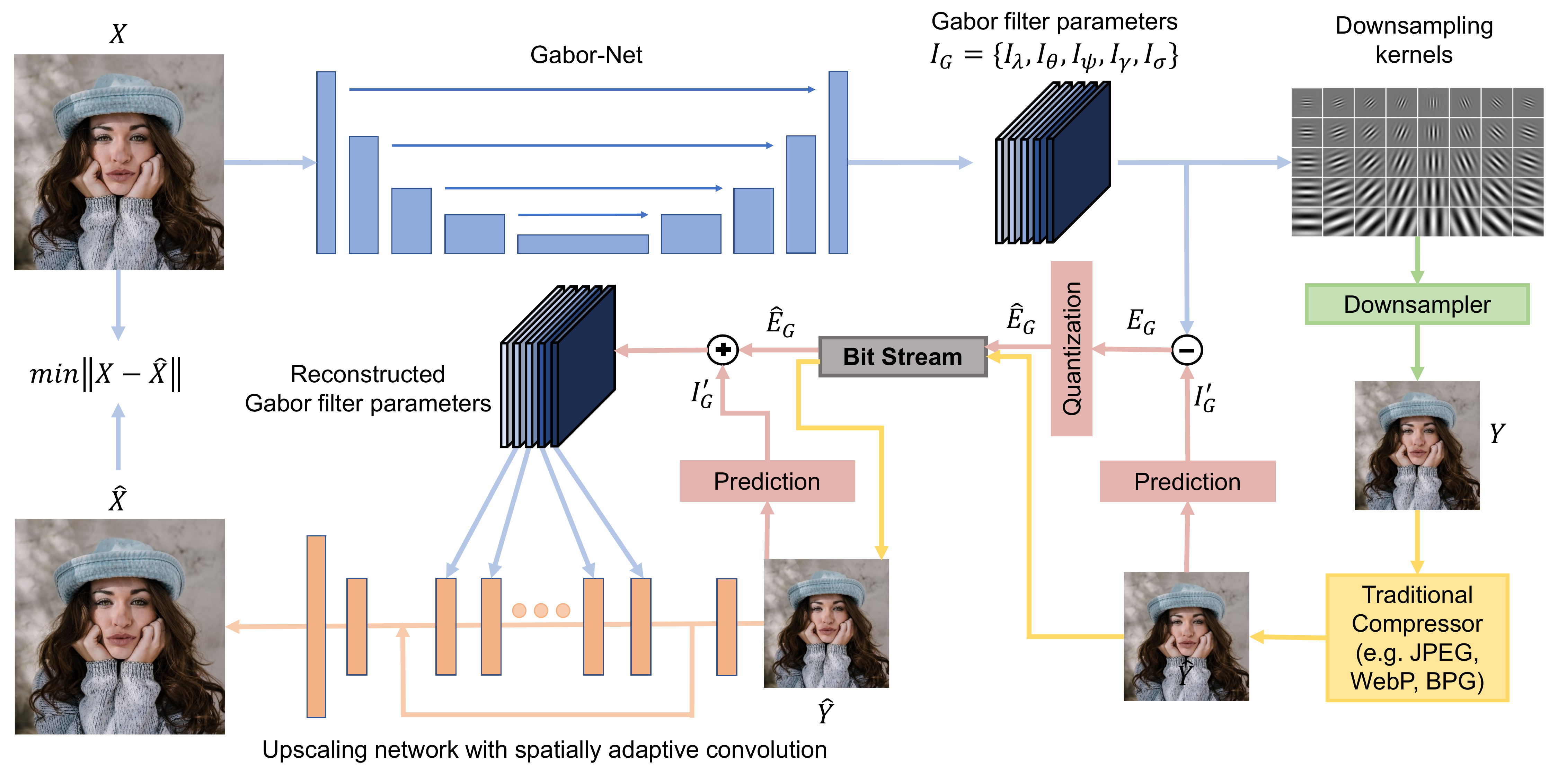}
	\caption{The overall framework of the proposed ADDL image compression system. It consists of a content-adaptive downsampling encoder and an spatially-varying upsampling decoder.}
	\label{addl}
\end{figure*}

\section{ADDL Compression System}
\label{sec:addl}
\subsection{Overview}
In this section, we design the proposed ADDL image compression system and present the following two
key technical developments, which are also the main contributions of this work.
1. learning content-adaptive downsampling kernels in the form of Gabor filters;
2. upconverting the downsampled and compressed image to its original resolution using a deep upsampling neural network, aided by the prior knowledge of the learned adaptive upsampling kernels.

The overall framework of the proposed ADDL image compression system is shown in Fig.~\ref{addl}.
The system consists of a content-adaptive downsampling encoder and an upsampling decoder.
Given an HR image $X$ to be compressed, the ADDL compression system first estimates the optimal downsampling kernels in each spatial location, then applies the estimated downsampling kernels to $X$ to produce the downsampled image $Y$. In our design, the downsampling kernels are restricted to the form of parametric Gabor filters to simplify the task and the architecture of the downsampling kernel optimization network. The parametric filter representation also reduces the side information needed to be transmitted to aid the upsampling at the decoder.
Next, the downsampled image $Y$ is compressed by any traditional image compressor (e.g. JPEG, JPEG200, WebP, BPG) to get the base layer $\hat{Y}$ for transmission and storage.
To achieve better upconversion result, the ADDL system not only transmits the base layer (downsampled and compressed image) $\hat{Y}$, but also the learned Gabor filter parameters.
The upsampling decoder takes the base layer (compressed downsampled image) $\hat{Y}$ and the Gabor filter parameters as input to produce the reconstructed HR image $\hat{X}$ by a network designed for joint super resolution and compression artifact reduction.

As illustrated by Fig.~\ref{addl}, the encoder and decoder networks, together with the spatially varying Gabor filters, are jointly optimized, via an end-to-end deep learning process, for the objective of minimizing the final reconstruction error $\| X-\hat{X} \|$.

%NOTE: minimizing the final HR error leads to optimal LR image?????

%Specifically, we design a deep neural network called Gabor-Net to estimate the optimal Gabor filter parameters $\{\lambda, \theta, \psi, \sigma, \gamma\}$ in each spatial location,
%
%then calculate the Gabor downsampling kernels based on the estimated parameters and downsampled image $X$ to get the downsampled Image $Y$.

\subsection{Content adaptive downsampling}
Traditional downsampling methods such as bilinear or bicubic, all have fixed downsampling kernels.
This content-independent approach is obviously not optimal and prone to aliasing artifacts.
In the ADDL system, spatially adaptive downsampling kernels are used for maximum information preservation.
The downsamplings kernel is optimized for each pixel using the corresponding context in the HR image.
Different from the existing method~\cite{sun2020learned} that directly optimizes the weights of the downsampling kernels, we restrict the downsampling kernels to the form of Gabor filters and optimize only the parameters. In the spatial domain, a 2-D Gabor filter is a Gaussian kernel function modulated by a sinusoidal plane wave:
\begin{equation}
\begin{split}
G  (x, y; \lambda, \theta, \psi, \sigma, \gamma) & =
    e^{-\frac{x'^2 + \gamma^{2}y'^2} {2\sigma^2}} cos(\frac{2\pi}{\lambda} x' + \psi) \\
    x' & = x cos\theta + y sin\theta \\
    y' & = -x sin\theta + y cos \theta
\end{split}
\label{eq:gabor}
\end{equation}
where $\lambda$ represents the wavelength of the sinusoidal factor, $\theta$ represents the orientation of the normal to the parallel stripes of a Gabor function, $\psi$ is the phase offset, $\sigma$ is the sigma/standard deviation of the Gaussian envelope and  $\gamma$ is the spatial aspect ratio, and specifies the ellipticity of the support of the Gabor function.
%human vision

The parametric representation of the Gabor filter reduces the problem dimensionality to five from  the size of convolution kernel, and hence can lead to a simpler CNN model, called Gabor-Net, for the task of optimizing downsampling filter kernels.  Another reason for our choice of the filter type is that the frequency and orientation characteristics of Gabor filters are similar to those of the human visual system.
\cite{olshausen1996emergence}.
% Olshausen, B. A. & Field, D. J. (1996). "Emergence of simple-cell receptive-field properties by learning a sparse code for natural images". Nature. 381 (6583): 607–609.

\begin{figure*}[t]
    \centering
    \includegraphics[width=0.98\linewidth]{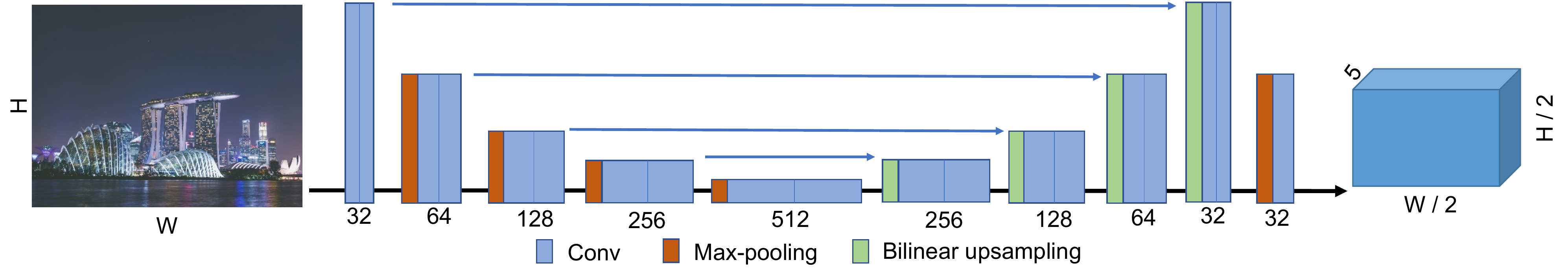}
    \caption{The architecture of the proposed Gabor-Net. It is a U-Net-like Encoder-Decoder network, trained to optimize the five parameters $\{\lambda, \theta, \psi, \sigma, \gamma\}$ of the Gabor filters for each pixel.}
    \label{gabornet}
\end{figure*}

The architecture of Gabor-Net $\mathcal{G}$ is shown in Fig.~\ref{gabornet}.
It is a U-Net-like Encoder-Decoder network, trained to optimize the five parameters $\{\lambda, \theta, \psi, \sigma, \gamma\}$ of the Gabor filters for each pixel.
The encoder part has an input convolution layer and five stages comprised of a max-pooling layer followed by two convolutional layers.
The input layer has 32 convolution filters with size of 3$\times$3 and stride of 1.
The first stage is size-invariant and the other four stages gradually reduce the feature map resolution by max-pooling to obtain a larger receptive field.
The decoder is almost symmetrical to the encoder. Each stage consists of a bilinear upsampling layer followed by two convolution layers  and a ReLU activation function. The input of each layer is the concatenated feature maps of the up-sampled output from its previous layer and its corresponding layer in the encoder.
For an 2-D input image $X$ of size $H*W$, since Gabor-Net learns an optimal Gabor filter at each position for downsampling, the output of Gabor-Net will be five 2-D maps $I_G = \{I_\lambda, I_\theta, I_\psi, I_\sigma, I_\gamma \}$ of size $5*\frac{H}{2}*\frac{W}{2}$, representing the five Gabor filter parameters, respectively.

The learned optimal Gabor filters are used to downsample the original image $X$ to produce the low-resolution image $Y$. The resulting low-resoution $Y$ can be further compressed by any traditional image compressor, e.g. JPEG, WebP, BPG, etc., and transmitted to the receiver.

\begin{figure}[t]
    \centering
    \includegraphics[width=0.98\linewidth]{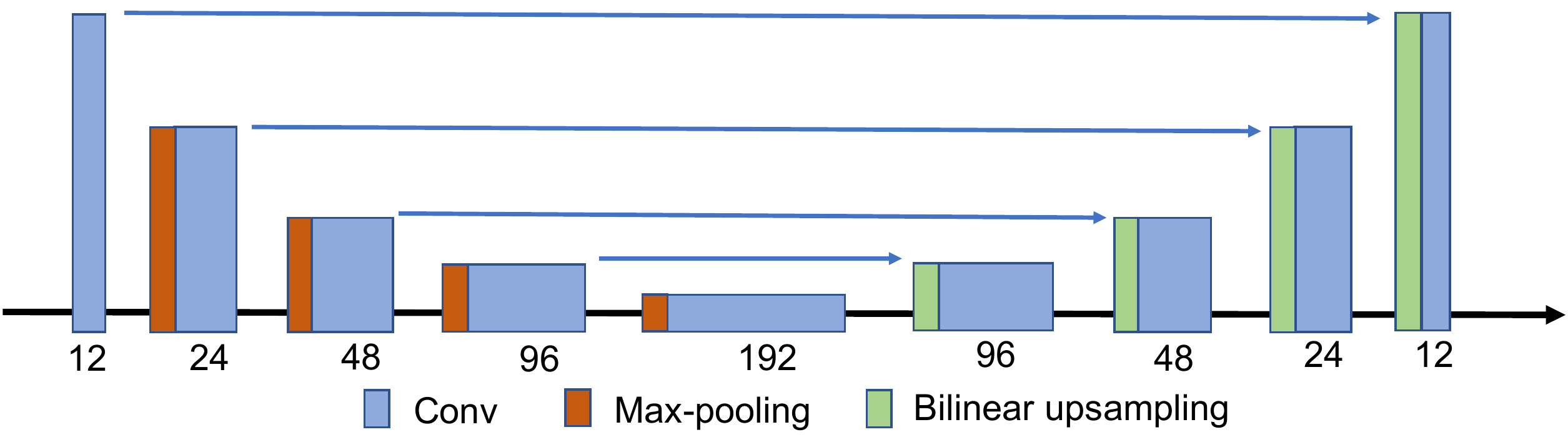}
    \caption{The architecture of the lightweight prediction network designed for predicting the Gabor filter parameters from the downsampled and compressed base layer image.}
    \label{prednet}
\end{figure}

\subsection{Predictive coding of Gabor parameters}
The learned downsampling kernels contain the information of the high-frequency content which is lost in the downsampling process, such as the orientations of the edges, so they are useful for reconstructing the high-frequency textures in the upconversion.
But this benefits compression only if we have a way of efficiently coding Gabor filter parameters.
%However, transmitting the downsampling kernels may leads to coding inefficiency, even though in our design we only need to transmit the Gabor filter parameters.
%For better coding efficiency,
Instead of directly transmitting the Gabor filter parameters $I_G$, we first predict the $I_G$ from the compressed low-resolution layer $\hat{Y}$ using a lightweight network, and further quantize the prediction residues for transmission.  The use of the predictive coding strategy ~\cite{calic} can greatly reduce the bit budget for transmitting the filter parameters while controlling the compression distortion to be below a threshold.

Specifically, we design a lightweight prediction network $\mathcal{L}$  (see Fig.~\ref{prednet}) to predict the Gabor filter parameters $I_G$ from $\hat{Y}$. The prediction network $\mathcal{L}$ has a similar architecture to the Gabor-Net, except that the number of convolution layers and convolution kernels per layer are decreased to reduce the complexity.
Let the output of the prediction network $\mathcal{L}$ be $I_G' = \mathcal{L}(\hat{Y})$.  The prediction residue $E_G = I_G - I_G'$ has a lower entropy than $I_G$. %Compared to the original parameters $I_G$, the entropy of the prediction residues $E_G$ is smaller than former.
We next quantize the prediction residues $E_G$ to further reduce the bit rates that need to be transmitted. Finally we only need to transmit the quantized prediction residues $\hat{E_G}$, that is:
\begin{equation}
    \hat{E_G} = \mathcal{Q}[ I_G - \mathcal{L}(\hat{Y}) ]
	\label{quan}
\end{equation}
where $\mathcal{Q}$ represents the quantization function. By adjusting the quantization step in $\mathcal{Q}$, we can control the bit rates of the transmitted quantized prediction residue to not exceed 20\% of the bit rates of the downsampled and coded base layer. 

\begin{figure}[t]
    \centering
    \includegraphics[width=0.98\linewidth]{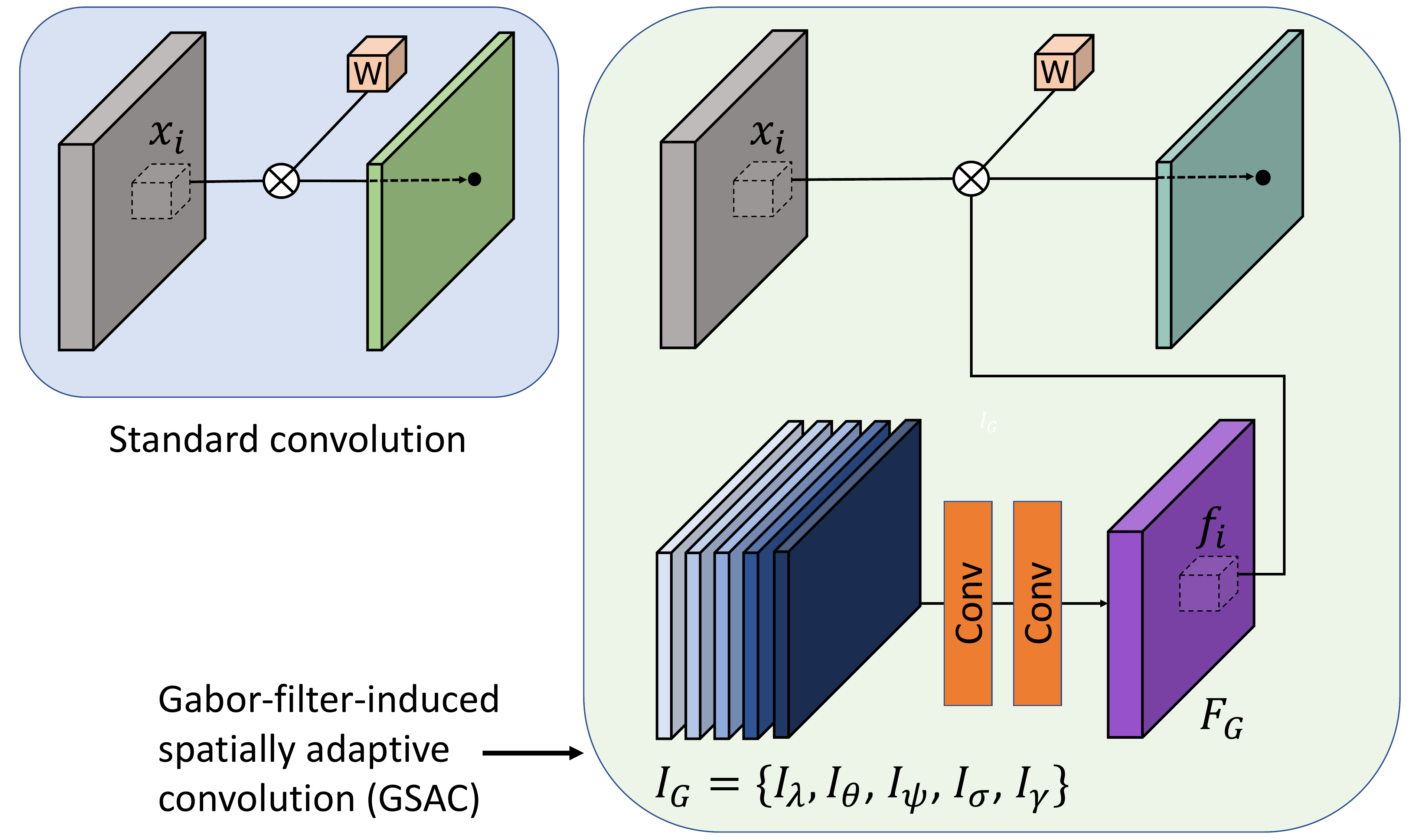}
    \caption{The Comparison between the standard convolution and the proposed Gabor-filter-induced spatially adaptive convolution.}
    \label{pac}
\end{figure}

\subsection{Upsampling with spatially adaptive convolution}
The upsampling decoder aims to upconvert the compressed downsampled image $\hat{Y}$ to the original resolution and reduce the compression artifacts. This task is very close to the existing image super-resolution technology~\cite{liu2020RFANet,zhang2020usrnet,liang21swinir,kai2021bsrgan,liang21manet,zhang2021DPIR}.
However, the existing super-resolution methods all adopt spatially invariant convolution as the basic unit to construct the deep neural network.  This is not the optimal choice for our task.
Instead, the ADDL decoder performs spatially varying upsampling to match the content-adaptive downsampling at the encoder.  It adopts pixel-adaptive convolution~\cite{pac} as the basic unit to construct the upsampling neural network.

%process is content adaptive, so the upsampling process should also be spatially variant. For this reason, we adopt the pixel-adaptive convolution~\cite{pac} as the basic unit to construct the upsampling neural network in our decoder.

A standard convolution operation can be defined as:
\begin{equation}
    y_i = W \cdot x_i + b
\end{equation}
where $x_i$ is the $i$-th convolution window, $W$ and $b$ are the convolution weight and bias.
Departing from the tradition, we propose a Gabor-filter-induced spatially adaptive convolution (GSAC), as shown in Fig.~\ref{pac}.
In GSAC, the Gabor filter parameters $I_G$ are fed into two convolutional layers to extract features $F_G$, then the resulting $F_G$ are used to modulate the standard convolution to make it spatially adaptive. This process can be formulated as:
\begin{equation}
    y'_i = (W \cdot f_i) \cdot x_i + b
\end{equation}
where $f_i$ represents the $i$-th convolution window in $F_G$.

\begin{figure}[t]
\centering
\includegraphics[width=0.98\linewidth]{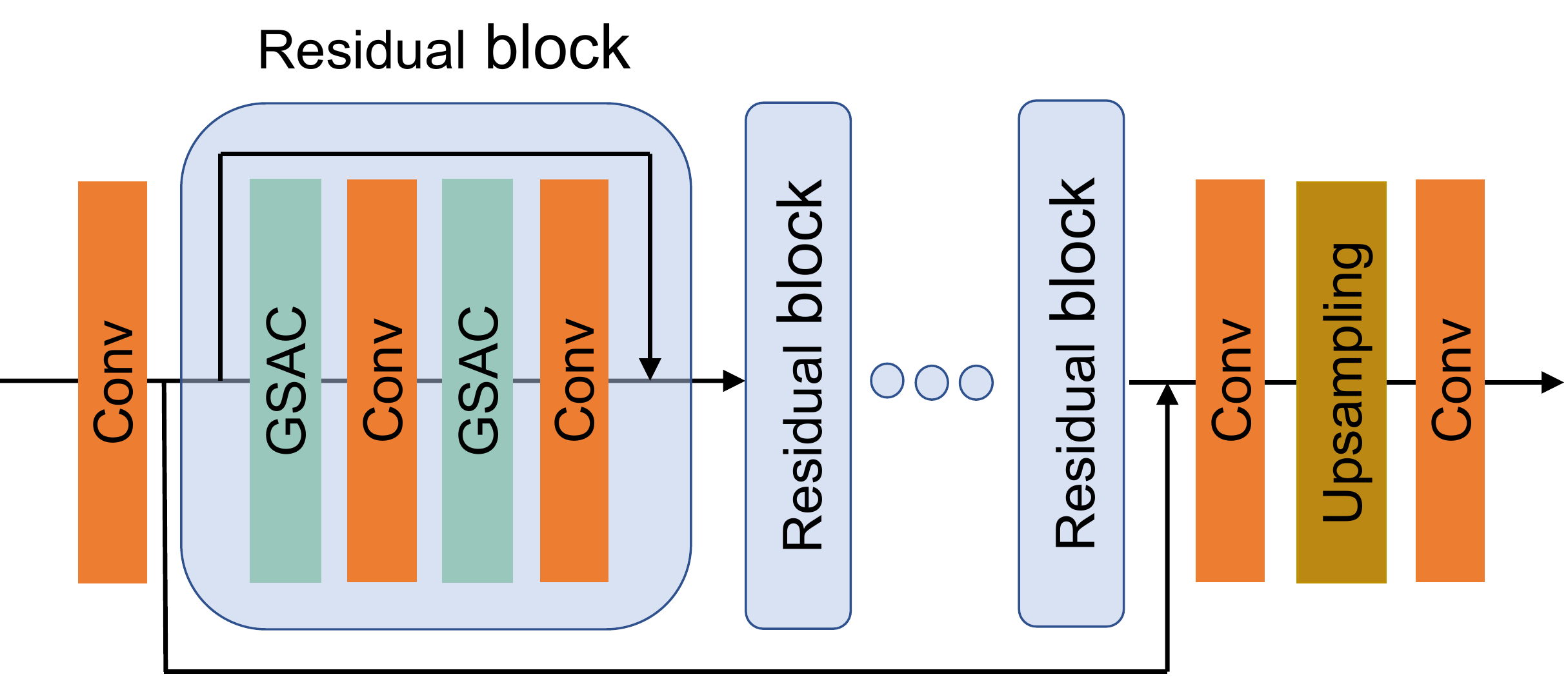}
\caption{Architecure of the proposed upsampling neural network.}
\label{upsampling}
\end{figure}

We design an upsampling network with the proposed GSAC method, as shown in Fig.~\ref{upsampling}.
It consists of 16 residual blocks with GSAC, in each of which there are two GSAC layers and two standard convolution layers. To avoid interference of different textures in one image, we restrict the receptive field of the GSAC by using $1 \times 1$ convolution kernels.
The standard convolution layers all adopt $3 \times 3$ convolution kernels.
Skip connection is used to ease the training of deep CNN.
The upsampling layer is implemented by the transposed convolution.

Now we are at the point to summarize the overall pipeline of ADDL compression system in Algorithm.~\ref{alg_agdl}.
\begin{algorithm}[!t]
\caption{ Framework of ADDL compression system.}
\label{alg_agdl}
\hspace*{0.02in} {\bf Input:}
The original image, $X$; \\
\hspace*{0.02in} {\bf Output:}
The decoded image, $\hat{X}$; \\
\hspace*{0.02in} {\bf Encoding:}
\begin{algorithmic}[1]
\STATE Learn content adaptive Gabor parameters $I_G$ from $X$;
\STATE Calculate the Gabor downsampling kernels according to the learned parameters $I_G$;
\STATE Downsample $X$ using the Gabor filter kernels to produce the low-resolution $Y$;
\STATE Compress $Y$ using a traditional image compressor (e.g. JPEG) to $\hat{Y}$;
\STATE Predict $I_G'$ from $\hat{Y}$ and quantize the prediction residues to get $\hat{E_G} = \mathcal{Q}[ I_G - I_G']$;
\STATE Transmit $\hat{Y}$ and $\hat{E_G}$.
\end{algorithmic}
\hspace*{0.02in} {\bf Decoding:}
\begin{algorithmic}[1]
\STATE Decode $\hat{Y}$ and $\hat{E_G}$ from the bit stream;
\STATE Predict $I_G'$ from $\hat{Y}$ and reconstruct the Gabor filter parameters by $\hat{I_G} = I_G'+\hat{E_G}$;
\STATE Construct the GSAC layer using $\hat{I_G}$ and nuild the upsampling network;
\STATE Reconstruct the high-resolution image $\hat{I}$.
\end{algorithmic}
\end{algorithm}

% \subsection{Network training}

\begin{figure*}[t]
	\centering
	\includegraphics[width=0.49\linewidth]{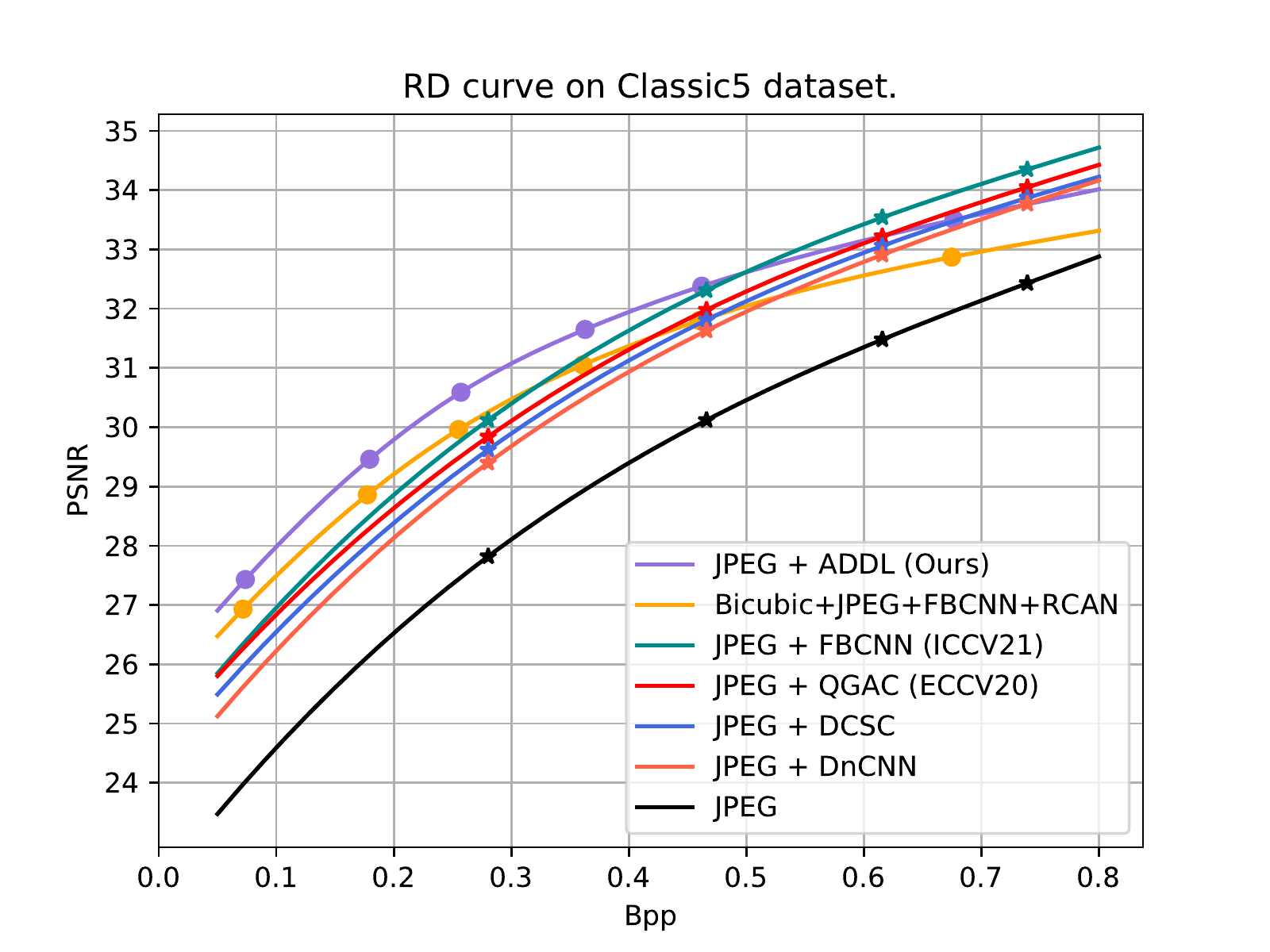}
	\hfill
	\includegraphics[width=0.49\linewidth]{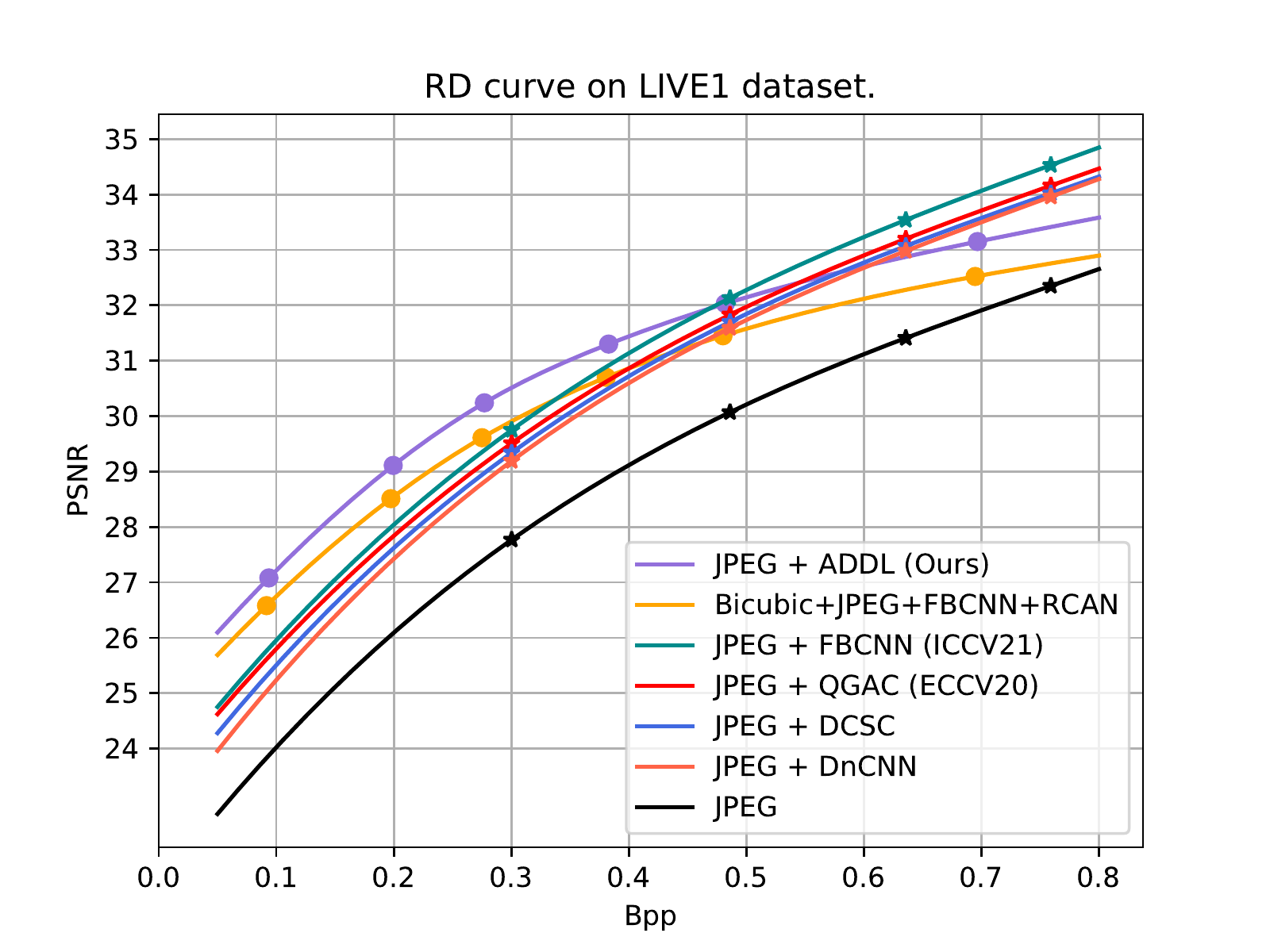} \\
	\includegraphics[width=0.49\linewidth]{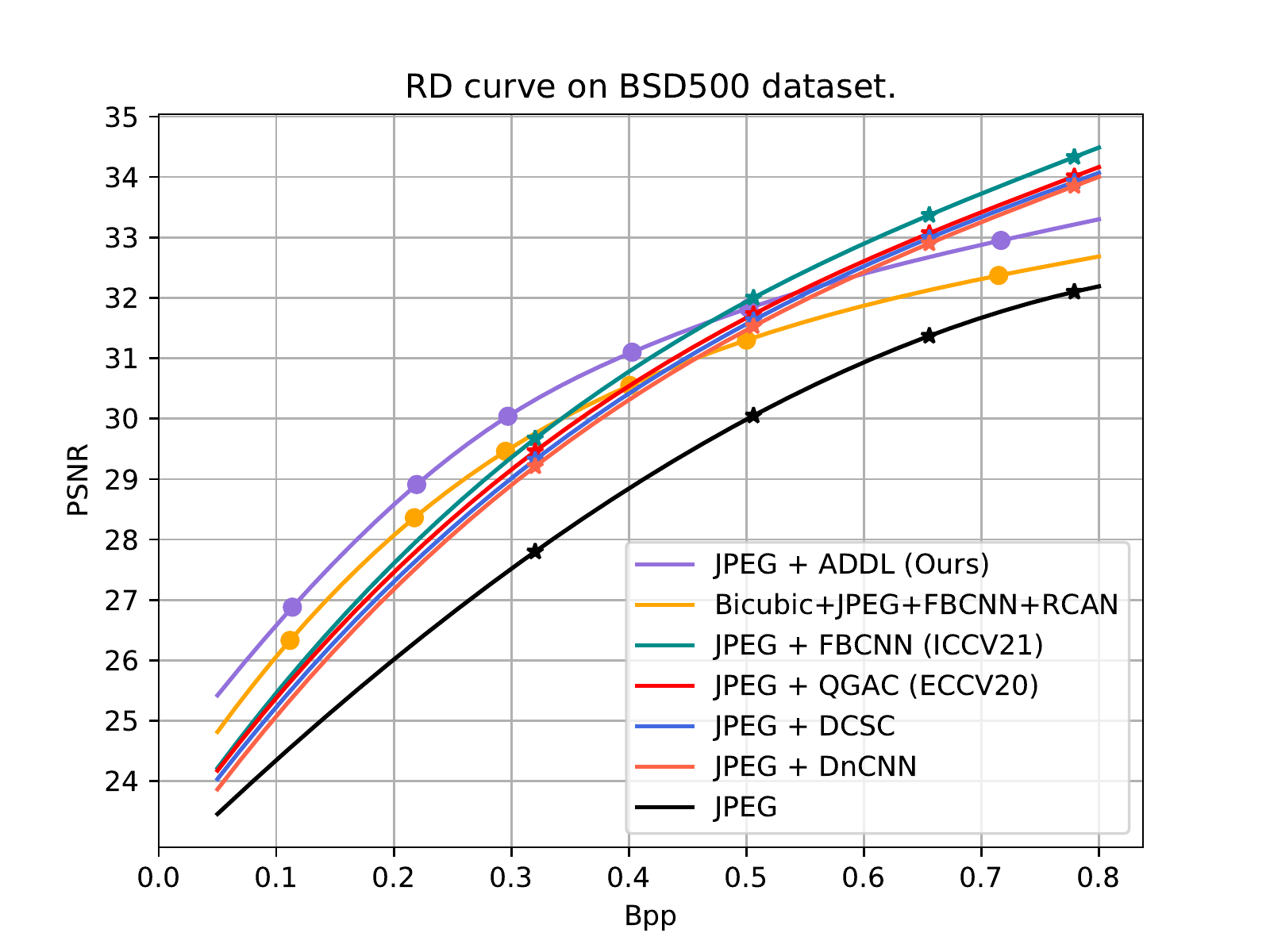}
	\hfill
	\includegraphics[width=0.49\linewidth]{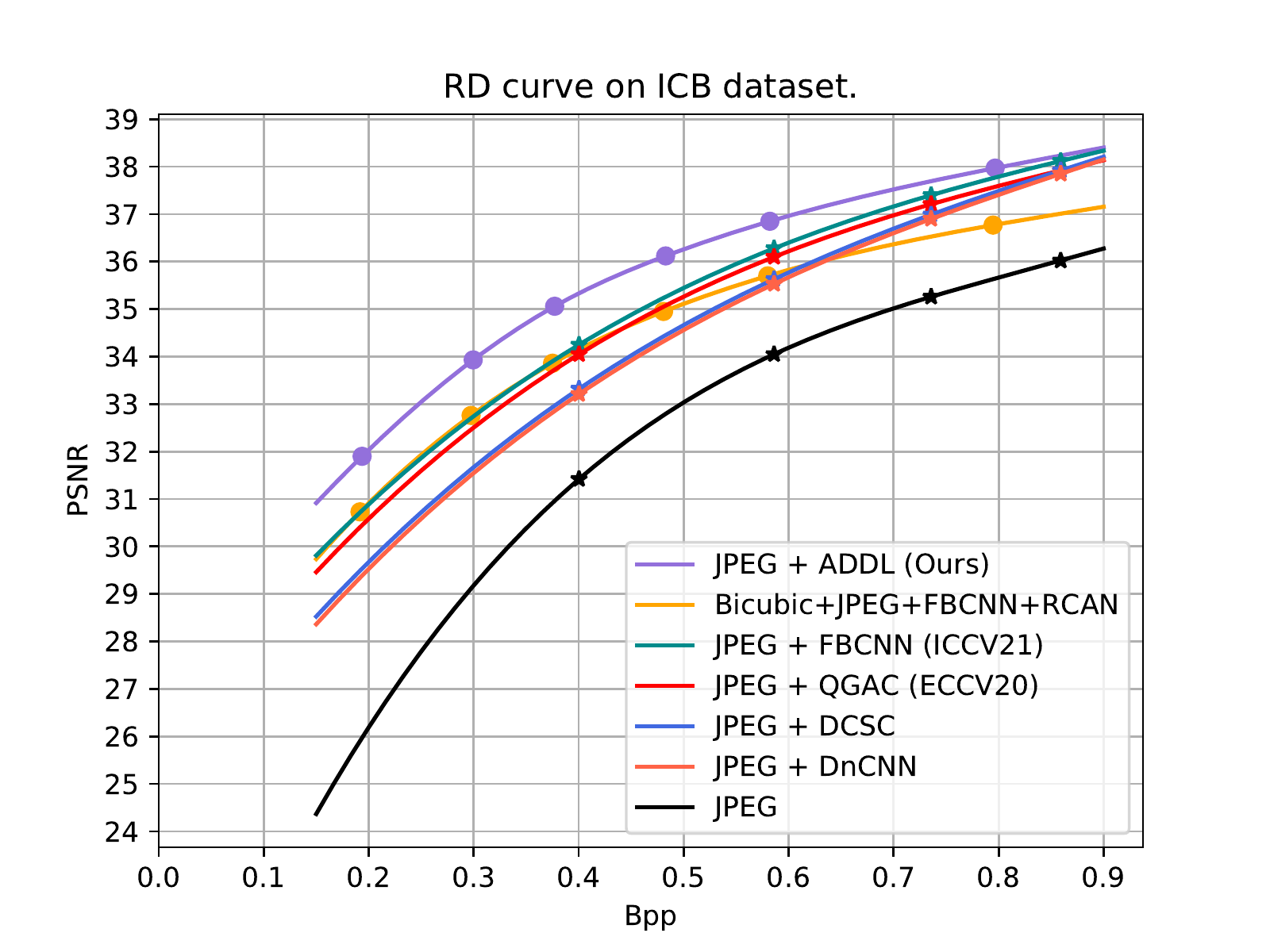}
	\caption{RD curves of JPEG + ADDL scheme and other competing methods on Classic5, LIVE1, BSD500 and ICB datasets.}
	\label{fig:rd_jpeg}		
\end{figure*}

\section{Experiments}
\label{sec:exps}
In this section, we present the implementation details of the proposed ADDL image compression system.  To systematically evaluate and analyze the performance of the ADDL compression system, we conduct extensive experiments and compare our results with several stat-of-the-art methods.

\subsection{Data preparation and network training}
For training the proposed ADDL compression system,
we use the widely-used high-quality 2K-resolution image dataset DIV2K~\cite{DIV2K} as our training data.
The set consists of 800 training images and 100 validation images.
For testing, we evaluate the trained model on the four commonly used benchmarks: Classic5~\cite{classic5}, LIVE1~\cite{LIVE1}, BSD500~\cite{BSD100} and ICB~\cite{icb}, and report the performances.

The whole pipeline of the proposed ADDL system is too complicated to do the end-to-end training. For this reason, we first train the Gabor-Net and the upsampling module without the side information of Gabor filter parameters via an end-to-end manner. After that, we train the prediction network $\mathcal{L}$ to predict the Gabor filter parameters from the downsampled and JPEG-compressed images. Finally, we finetune the upsampling network by incorporating the transmitted Gabor filter parameters.

During training, we randomly extract patches of size $256 \times 256$.
All training processes use the Adam~\cite{adam} optimizer by setting $\beta_1 = 0.9$ and $\beta_2 = 0.999$, with a batch size of $64$.
The learning rate starts from $1 \times 10^{-4}$ and decays by a factor of 0.5 every $4 \times 10^{4}$ iterations and finally ends with $1.25 \times 10^{-5}$. 
$\mathcal{L}_1$ loss is adopted to optimize all networks in the ADDL compression system.
We train our model with PyTorch on four NVIDIA GeForce GTX 2080Ti GPUs. It takes about two days to converge.
All the training and evaluation processes are performed on the luminance channel (in YCbCr color space).

We choose JPEG as the traditional image compressor in ADDL as it is by far the most common image compression method. During training the JPEG quality factor is randomly sampled from 10 to 90.
However, JPEG compression algorithm contains quantization/rounding operation, and the rounding operation $\lfloor \cdot \rceil$ has derivative 0 nearly everywhere, which is not compatible with the gradient-based optimization, so it can not be directly embedded into the training process.
To solve this problem, following the solution in~\cite{shin2017jpeg},
we instead use the approximation
$\lfloor x \rceil _{approx}= \lfloor x \rceil + (x - \lfloor x \rceil)^{3}$,
which has non-zero derivatives nearly everywhere, and close to $\lfloor x \rceil$.

we also build ADDL system with other traditional image compressors (such as BPG) and report the experimental results in the following subsections.

\begin{figure*}[t]
	\centering
	\includegraphics[width=0.88\linewidth]{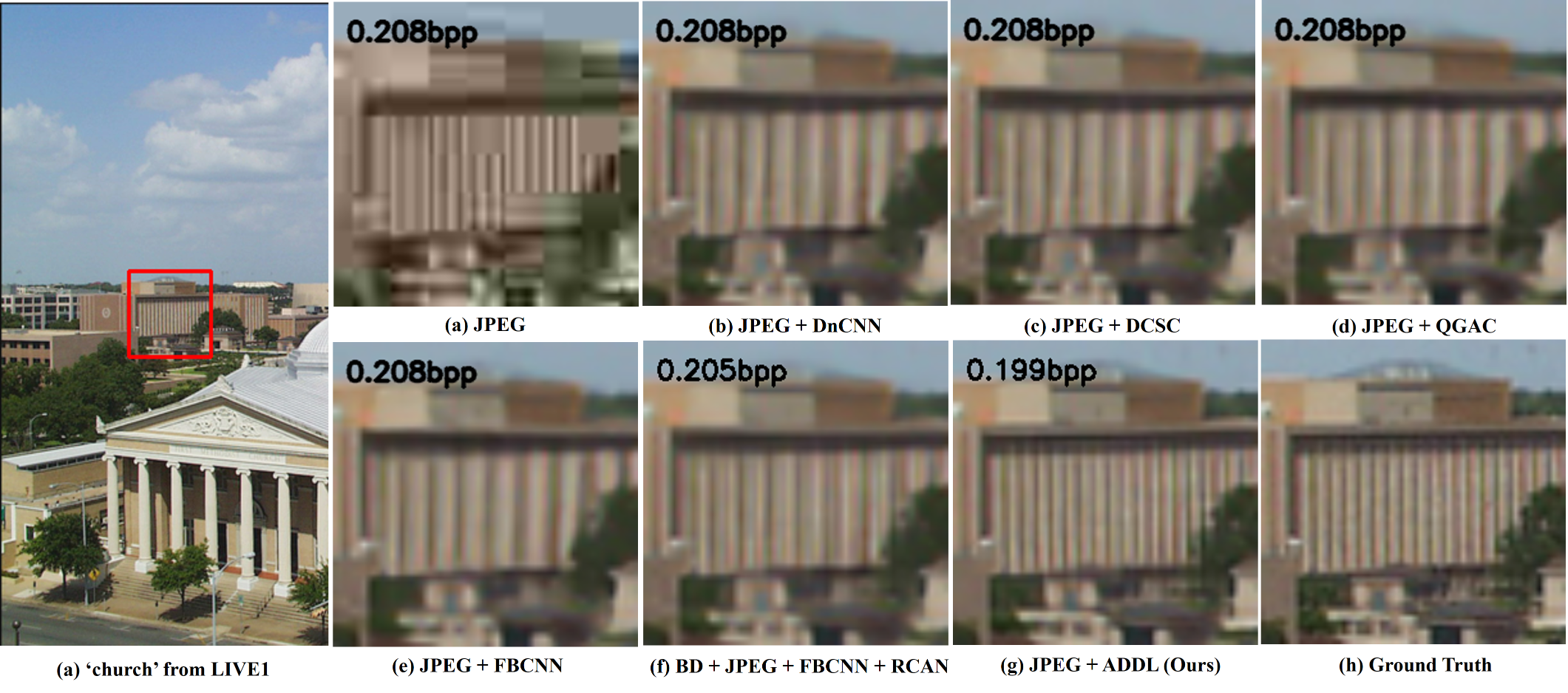}
        \caption{Visual comparisons of JPEG + ADDL scheme and other competing methods on the Image 'LIVE1: church'. BD means bicubic downsampling.}
        \label{church}		
\end{figure*}
\begin{figure*}[t]
        \centering
	\includegraphics[width=0.88\linewidth]{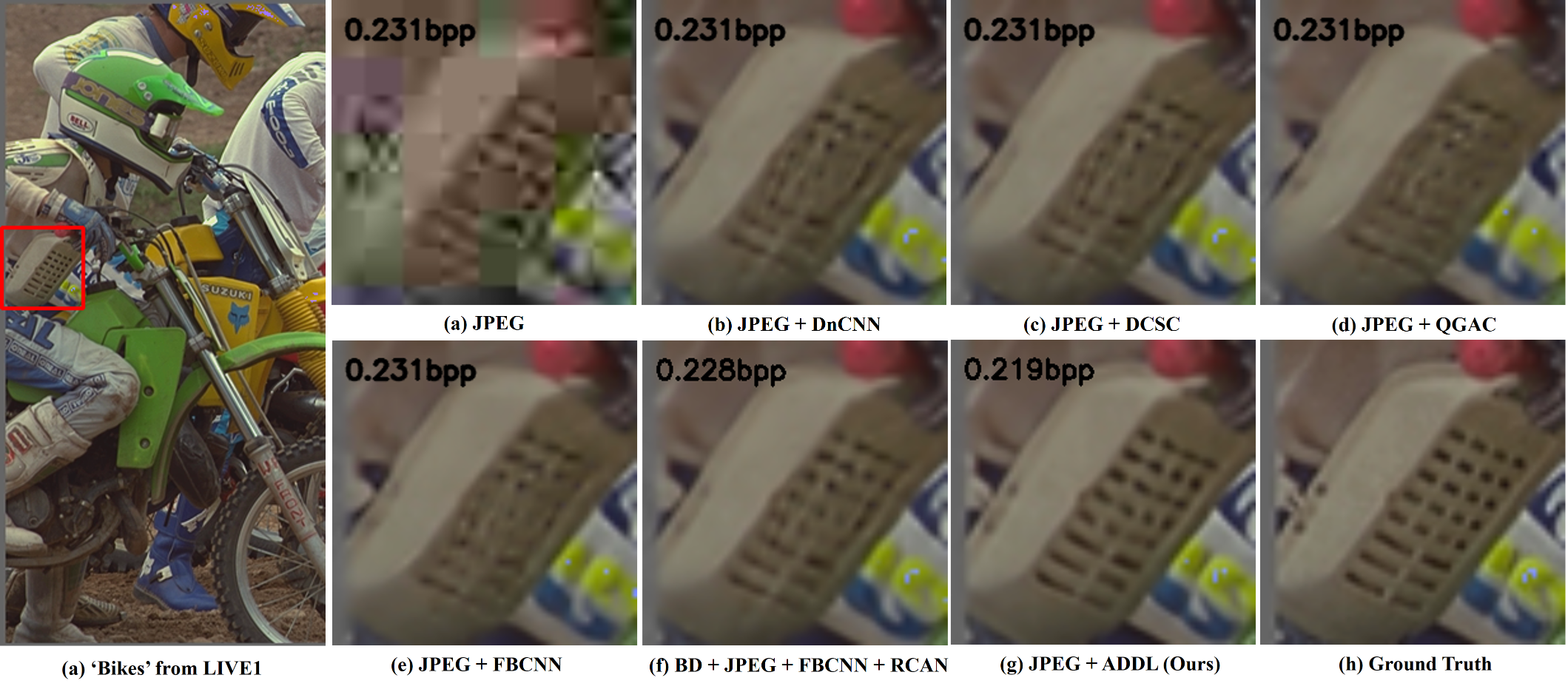}
	\caption{Visual comparisons of JPEG + ADDL scheme and other competing methods on the Image 'LIVE1: bikes'. BD means bicubic downsampling..}
	\label{bikes}		
\end{figure*}

\begin{figure*}[t]
	\centering
	\includegraphics[width=0.88\linewidth]{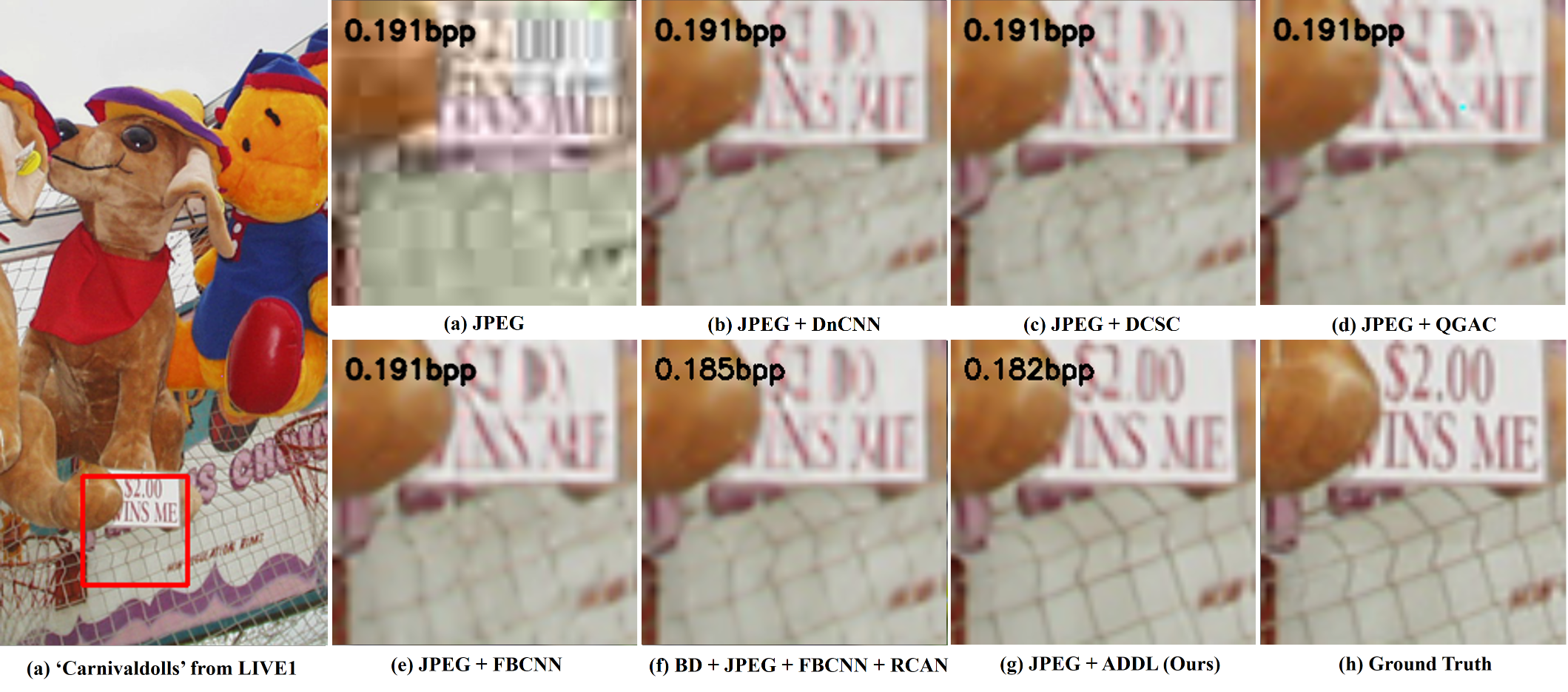}
	\caption{Visual comparisons of JPEG + ADDL scheme and other competing methods on the Image 'LIVE1: carnivaldolls'. BD means bicubic downsampling.}
	\label{carnivaldolls}		
\end{figure*}
\begin{figure*}[!h]
	\centering
	\includegraphics[width=0.88\linewidth]{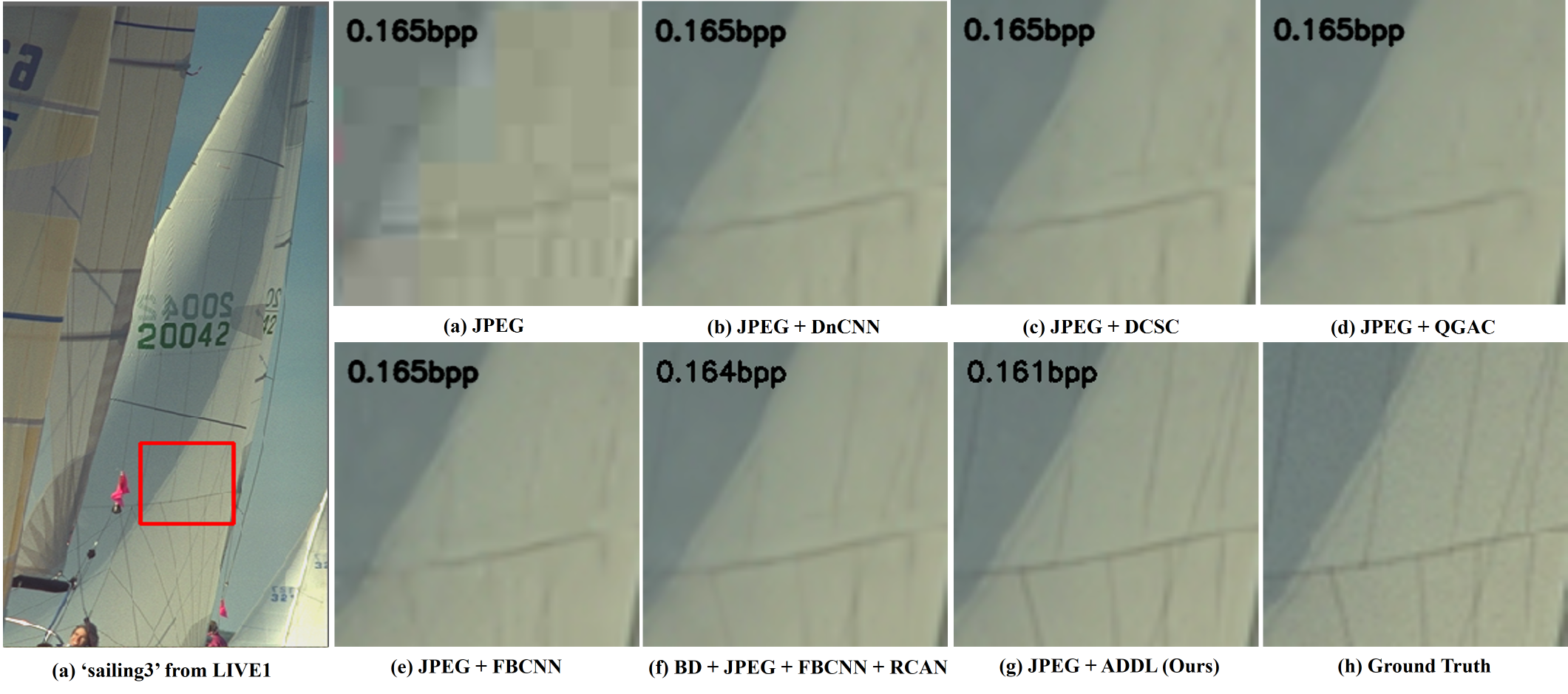}
	\caption{Visual comparisons of JPEG + ADDL scheme and other competing methods on the Image 'LIVE1: sailing3'. BD means bicubic downsampling.}
	\label{sailing3}		
\end{figure*}

\subsection{ADDL with JPEG}
To demonstrate the advantages of the proposed ADDL compression system, we compare ADDL with several other compression systems, in which JPEG is also used as the image compressor and different learning-based post-processing algorithms (super-resolution, compression artifact reduction) are used for restoring/enhancing the compressed images.
We divide these competing compression systems into two categories:\\
\textbf{1. JPEG + Deblocking}.
In this category, several learning-based deblocking (also called compression artifact reduction) methods: DNCNN~\cite{dncnn}, DCSC~\cite{dcsc}, QGAC~\cite{qgac}, FBCNN~\cite{fbcnn} are combinded with JPEG compressor to form the competing image compression systems.\\
\textbf{2. Downsampling + JPEG + Deblocking + SR}.
This competing system consists of: downsampling by the fixed kernels, traditional compression, joint CNN-based deblocking and super-resolution.
we choose bicubic downsampling, FBCNN~\cite{fbcnn} and RCAN~\cite{rcan} as downsampling, deblocking and super-resolution method, respectively.

In the ADDL system, the total bit rates need to be transmitted are the sum of the rates of JPEG-coded low-resolution layer and the quantized prediction residues of the learned content adaptive Gabor filter parameters. To facilitate fair rate-distortion performance evaluations, for each test image, the rates of the competing compression systems are adjusted to match or be slightly higher than that of the ADDL compression system.
It is noteworthy that by adjusting the quantization step of $\mathcal{Q}$ in Eq.~\ref{quan}, for different bit rates (or compression ratios), the bit rates of the quantized prediction residues of the Gabor filter parameters are controlled to be about 20\% of the bit rates of the JPEG-coded base layer. 

\textbf{Quantitative evaluation.}
We present rate-distortion (RD) curves of the competing methods in Fig.~\ref{fig:rd_jpeg}.  As shown in the figure, the proposed ADDL compression system outperforms all the competing image compression methods consistently in PSNR measure at low to medium bit rates.
For Classic5, LIVE1 and BSD500 datasets, the proposed ADDL achieves superior rate-distortion performances to the best of other methods (JPEG + FBCNN), for bit rates lower than 0.5bpp. For the ICB dataset, ADDL beats all other methods for bit rates lower than 0.9bpp.

\textbf{Perceptual quality comparison.}
The perceptual qualities of competing methods, given the same bit rate, are compared in Figs.~\ref{church}, ~\ref{bikes}, ~\ref{carnivaldolls} and ~\ref{sailing3}.
% ~\ref{carnivaldolls} and ~\ref{sailing3}.
It can be seen that ADDL preserves high-frequency textures, such as meshes and letters, much better than the other compression methods.  At modest bit rates, ADDL achieves visually transparent quality compared with the ground truth, while the other methods still suffer from highly noticeable distortions.  Both figures clearly demonstrate the advantage of the content-dependent downsampling of ADDL (exhibit (g)) over spatially-invariant bicubic downsampling (exhibit (f)).

%For instances, ADDL faithfully reconstructs the spokes of the wheel while the other methods fail to, erasing or distorting the spoke structures; in Fig.~\ref{carnivaldolls}, the letters restored by ADDL are much cleaner and sharper than other methods with a visually lossless quality compared with the ground truth.

\subsection{ADDL with BPG}
\begin{figure*}[t]
	\centering
	\includegraphics[width=0.49\linewidth]{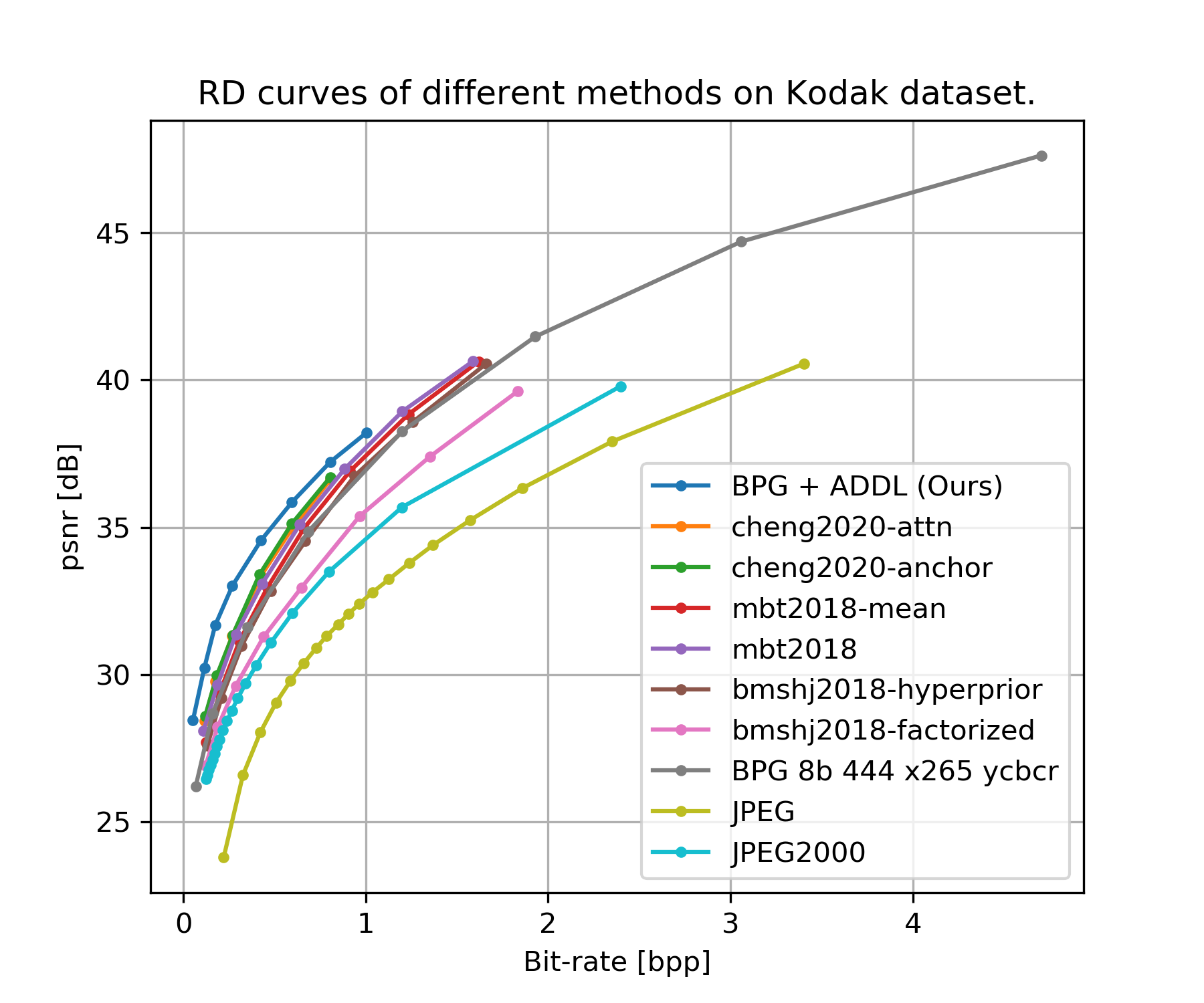}
    \hfill
	\includegraphics[width=0.49\linewidth]{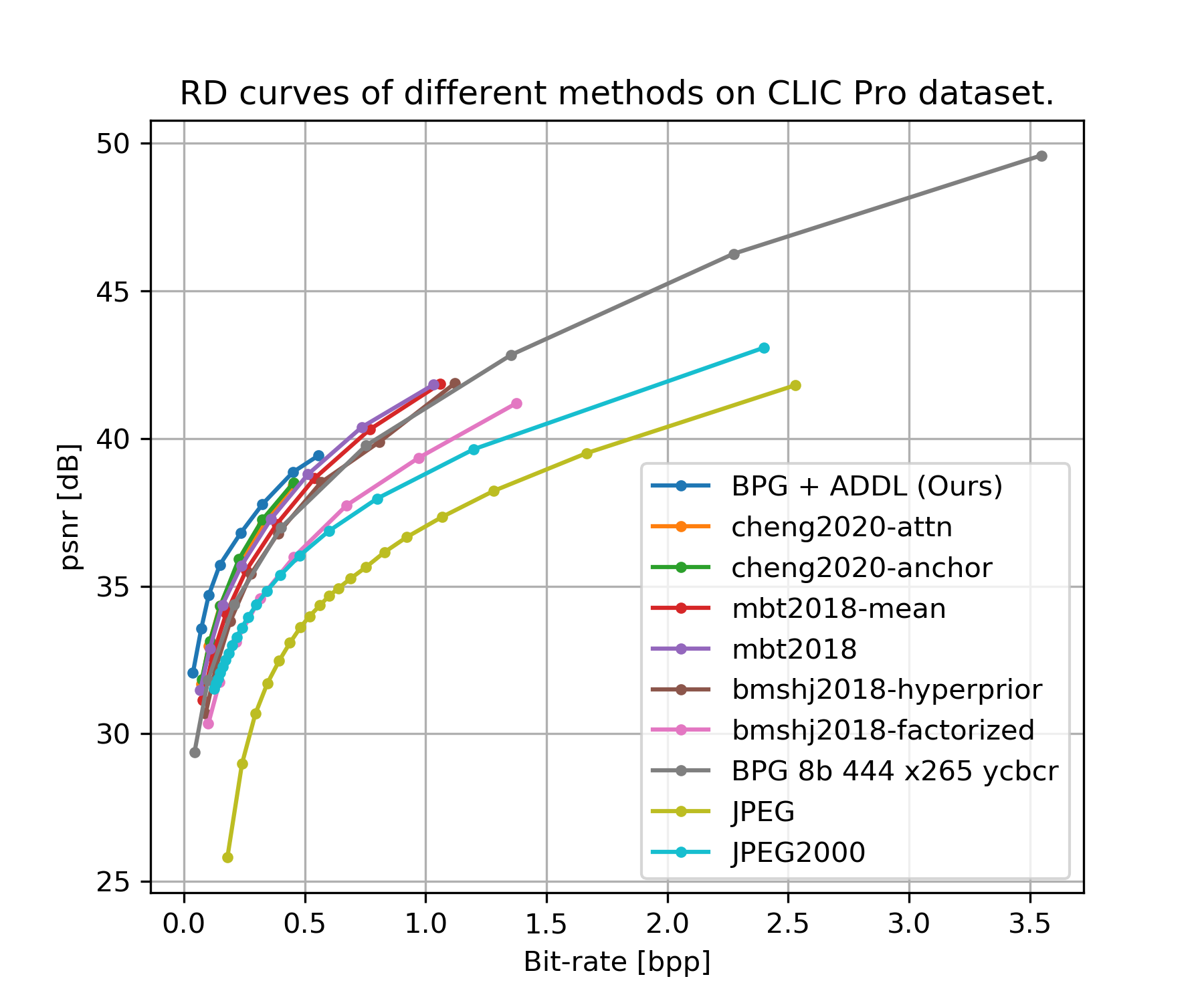}
	\caption{RD curves of BPG + ADDL scheme and other end-to-end optimized image compression methods on Kodak and CLIC Pro datasets.}
	\label{fig:rd_bpg}
\end{figure*}

In addition to JPEG, we also build ADDL system with the most powerful traditional image compressor BPG and compare the BPG + ADDL scheme with end-to-end optimized image compression methods.
We conduct experiments on Kodak and CLIC Pro datasets and present the reate-distortion curves in Fig.~\ref{fig:rd_bpg}. 
It can be seen that BPG + ADDL compression system outperforms all
the end-to-end optimized image compression methods consistently in PSNR measure at low bit rate level.

\subsection{Visualization of Gabor filter parameters}
\begin{figure}[t]
	\centering
	\includegraphics[width=0.95\linewidth]{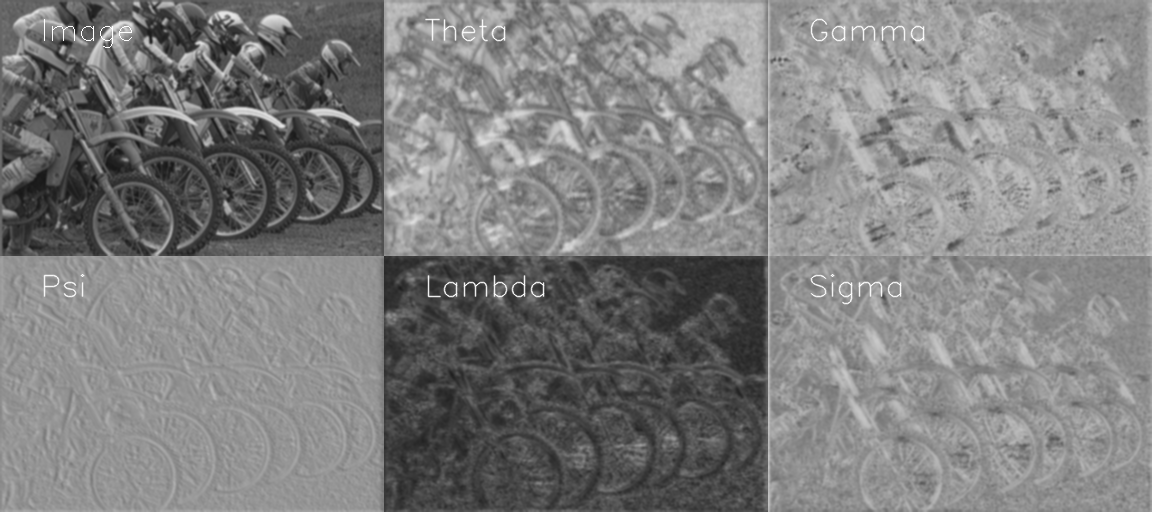}
	\caption{Visualization of the learned Gabor filter parameters.}
	\label{params_vis}		
\end{figure}

To have a more intuitive understanding of the learned Gabor downsampling kernels, let us visualize the learned Gabor filter parameters in Fig.~\ref{params_vis}.
It can be observed that the learned Gabor filter parameters are highly correlated to the image features and structures, especially for parameter $\theta$, which determines the orientation of the Gabor downsampling kernels.  This explains why spatially adaptive Gabor downsampling kernel
can preserve high-frequency information, and also why predictive coding of Gabor filter parameters works.

%observation implies that the proposed adaptive Gabor downsampling model will choose the downsampling kernel in the same orientation of the edge textures in the image, to preserve the

\begin{figure}[!h]
	\centering
	\includegraphics[width=0.95\linewidth]{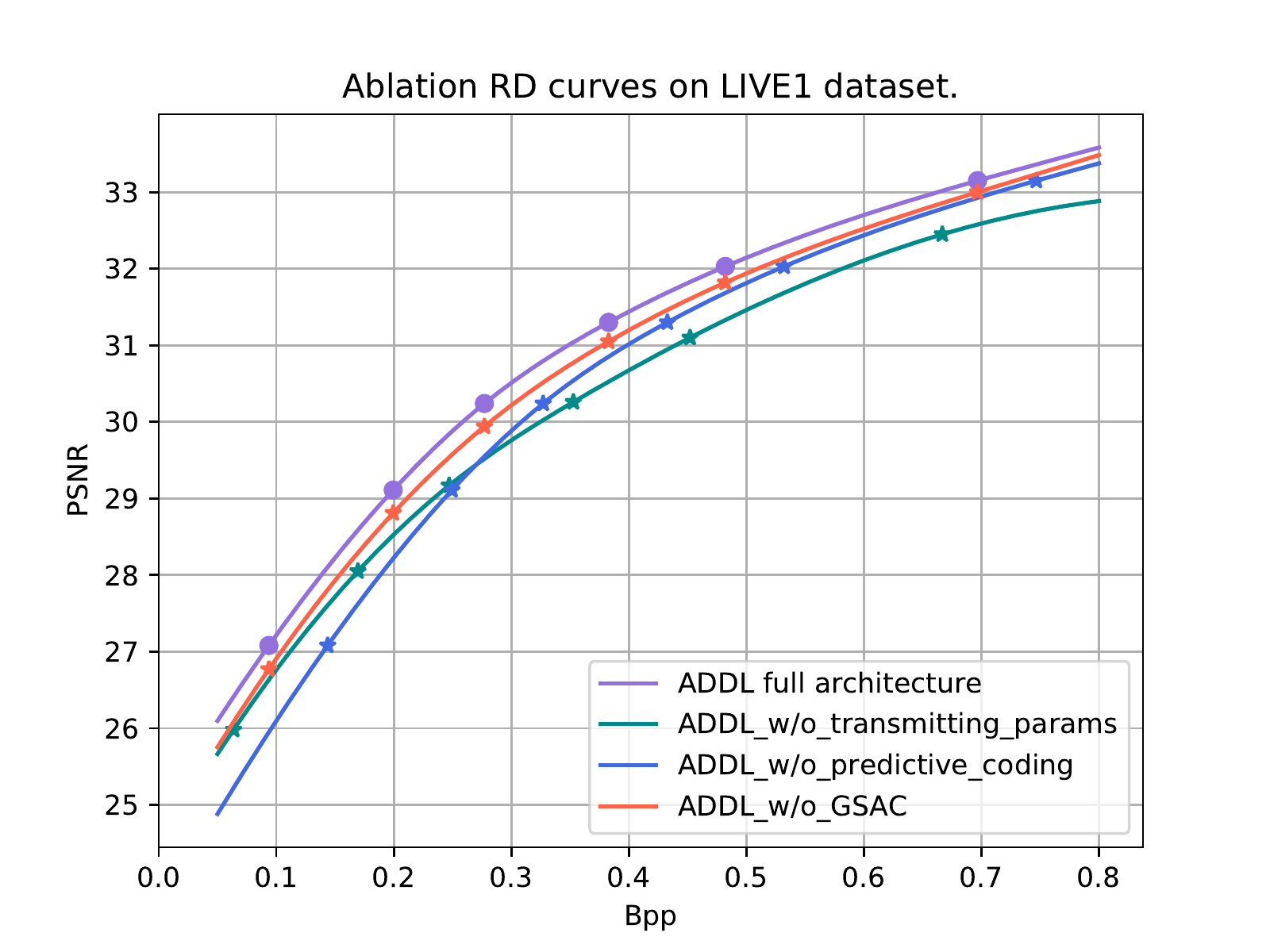}
	\caption{Ablation RD curves on LIVE1 dataset.}
	% \vspace{-0.1cm}
	\label{rd_ablation}		
\end{figure}

\subsection{Ablation Study}
In this subsection, we test various ablations of our full
architecture to evaluate the effects of each component of
the proposed ADDL compression system.

% \subsubsection{Ablation of transmitting Gabor filter parameters.}
We first evaluate the effects of whether to transmit the Gabor filter parameters and also how to transmit them to the receiver.
We build an ablation architecture which does not transmit the Gabor filter parameters at all (called ADDL\_w/o\_transmitting\_params) and another ablation architecture that directly transmits the Gabor filter parameters without predictive coding (called ADDL\_w/o\_predictive\_coding). The performances of these two architectures are shown in Fig.~\ref{rd_ablation}.
We can see that these two ablation architectures perform much worse than the full ADDL architecture in rate-distortion measure.

% \subsubsection{Ablation of GSAC module.}
We also build an another ablation network architecture (ADDL\_w/o\_GSAC), which transmits the Gabor filter parameters using the proposed predictive coding method, but the decoder uses only the received Gabor filter parameters without the GSAC module in the HR reconstruction.  We present the rate-distortion curve of this case in Fig.~\ref{rd_ablation}. It can be seen that without the GSAC module, the performance of the ADDL compression system drops a bit.
The decline in ADDL performance shows the effectiveness of the proposed GSAC module.

\section{Conclusion}
\label{sec:Conclusion}

We propose, implement and evaluate the new deep learning based ADDL image compression system.
The key idea is to code an image into a compact two-layer representation: a base layer that is generated by learned content-adaptive downsampling, and a refinement layer that is generated by a deep upsampling network.  
% The two layers are optimized jointly.  
The ADDL encoder and decoder collaborate through the sharing of information on spatially varying Gabor downsampling filters.

%%%%%%%%% REFERENCES

\bibliographystyle{IEEEtran}
\bibliography{addl}

\begin{IEEEbiography}
  [{\includegraphics[width=1in,clip,keepaspectratio]{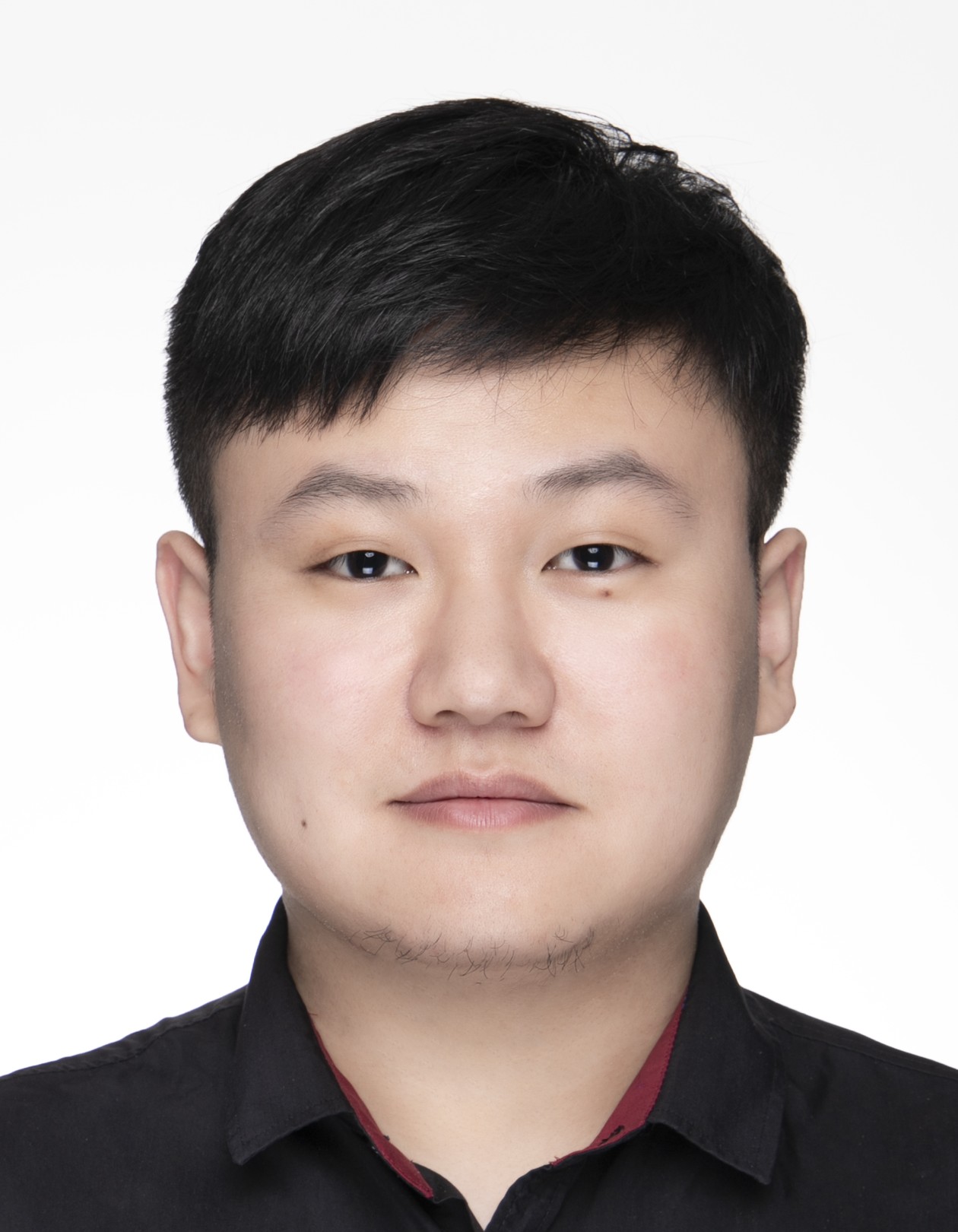}}]{Xi Zhang}
  received the B.Sc. degree in mathematics and physics basic science
  from University of Electronic Science and Technology of China, in 2015, and the Ph.D. degree 
  in electronic engineering from Shanghai Jiao Tong University, China, in 2022.
  He is currently a Postdoctoral Fellow with the Department of Electronic Engineering, 
  Shanghai Jiao Tong University, China.
  He was also a visiting Ph.D. student with the 
  Department of Electrical and Computer Engineering, 
  McMaster University, Hamilton, ON, Canada.
  His research interests include image and video processing, especially in image and video compression, enhancement, etc.
  He is also interested in other deep learning tasks such as transfer learning and visual reasoning.
\end{IEEEbiography}

\begin{IEEEbiography}
  [{\includegraphics[width=1in,clip,keepaspectratio]{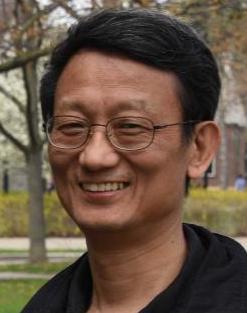}}]{Xiaolin Wu}
  (Fellow, IEEE) received the B.Sc. degree in computer science
  from Wuhan University, China, in 1982, and the Ph.D. degree in computer
  science from the University of Calgary, Canada, in 1988. He started his
  academic career in 1988. He was a Faculty Member with Western University,
  Canada, and New York Polytechnic University (NYU-Poly), USA. He is
  currently with McMaster University, Canada, where he is a Distinguished
  Engineering Professor and holds an NSERC Senior Industrial Research Chair.
  His research interests include image processing, data compression, 
  digital multimedia, low-level vision, and network-aware visual communication.
  He has authored or coauthored more than 300 research articles and holds
  four patents in these fields. He served on technical committees of many IEEE
  international conferences/workshops on image processing, multimedia, data
  compression, and information theory. He was a past Associated Editor of
  IEEE TRANSACTIONS ON MULTIMEDIA. He is also an Associated Editor of
  IEEE TRANSACTIONS ON IMAGE PROCESSING.
\end{IEEEbiography}

\end{document}